\documentstyle[12pt]{article}
\newcommand{\tr}{{\rm tr}}
\newcommand{\Q}{{\rm Q}}
\newcommand{\G}{{\rm G}}
\newcommand{\res}{{\rm res}}
\date{}
\begin{document}
\title{THE INTERACTION FORCE BETWEEN ROTATING BLACK HOLES AT EQUILIBRIUM
\footnote{journal reference:Theor. Math. Phys.
v.116, p.1024(1998)}}
\author{G. G. Varzugin}
\maketitle
\begin{center}{\it
Laboratory of Complex System Theory, Physics Institute of
St-Petersburg University, St. Petersburg, Peterhof, Ulyanovskaya
1, 198904. E-mail varzugin@paloma.spbu.ru}\end{center} \vskip1cm

\begin{abstract}
We study the previously constructed Riemann problem whose solutions
correspond to equilibrium configurations of black holes. We evaluate
the metric coefficients at the symmetry axis and the interaction force
between the black holes.
\end{abstract}

\section {Introduction}

In the previous work \cite{Varzugin97}, we studied the axially
symmetric stationary vacuum solution of Einstein equations
describing the equilibrium configuration of rotating black holes.
The equilibrium configuration was understood to mean a stationary
solution possessing a disconnected event horizon. All such
solutions satisfy a certain boundary-value problem for a system of
elliptic nonlinear equations that can be easily obtained from
\cite{Carter75}, where the regularity conditions for the symmetry
axis and the event horizon were first formulated. Our
investigation showed that this boundary-value problem reduces to a
matrix Riemann problem with a rational conjugation matrix. We also
noted that solutions corresponding to stationary equilibrium
configuration of black holes are likely to develop a conical
singularity on the symmetry axis, which is the reason why the
black hole do not fall on each other. We note that for rotating
black holes, the presence of conical singularity for arbitrary
values of the solution parameters is not quite obvious from the
physical standpoint, because it can be assumed that the rotation
of black holes leads to repulsion forces that can compensate the
gravitational attraction.

The conical singularity itself allows us to introduce the notion of the
interaction force between the black holes. We consider this in more
detail in the next section. The analytic expression for the interaction
force between Schwarzschild black holes has long been known \cite{Weyl21}.
The main result of the present work is our generalization of this to
the case of rotating black holes.

\section{The boundary-value problem}

In cylindrical coordinates, the axially symmetric metric is
$$ds^2=-Vdt^2+2Wdtd\phi+Xd\phi^2+{X\over\rho^2}e^{\beta}(d\rho^2+dz^2),$$
where the metric coefficient depend only on $\rho$ and $z$. The
Einstein equations are then divided into two system of nonlinear
equations, the first of which can be brought to the form $$(\rho
g_{,\rho}g^{-1})_{,\rho}+(\rho g_{,z}g^{-1})_{,z}=0,
\;\;g=\pmatrix{-V&W\cr W&X\cr},\;\;\det g=-\rho^2,\eqno(2.1)$$ and
the second one to the form $$ (\ln
({X\over\rho^2}e^{\beta}))_{,\zeta}={i\over 2\rho}\left(1+\tr(\rho
g_{,\zeta}g^{-1})^2\right),\;\;\partial_\zeta={1\over 2}
(\partial_z-i\partial_\rho).\eqno(2.2)$$

Let the event horizon have $N$ disconnected components, let
$z_1,\ldots z_{2N}$ be the $z$-coordinates of the intersection
points of the event horizon and the symmetry axis, and let
$\Omega_i$ be the angular velocity of the $i$th black hole. In the
neighborhood of the $i$th black hole, we choose the coordinate
system $$\rho^2=(\lambda^2-m_i^2)(1-\mu^2),\;\;
m_i={z_{2i}-z_{2i-1}\over2},$$
$$z-{z_{2i}+z_{2i-1}\over2}=\lambda\mu,\;\;|\mu|\leq1,$$ in which
the regularity condition of the symmetry axis and the event
horizon can be written as \cite{Carter75}
$$\pmatrix{1&\Omega_i\cr0&1\cr}g\pmatrix{1&0\cr\Omega_i&1\cr}=
\pmatrix{(\lambda^2-m_i^2)\hat V(\lambda,\mu)&\rho^2\hat
W(\lambda,\mu)\cr \rho^2\hat W(\lambda,\mu)& (1-\mu^2)\hat
X(\lambda,\mu)\cr},\eqno(2.3)$$ where $\hat X$ and $\hat V$ are
smooth functions not equal to zero. Using (2.3), we can easily
obtain the boundary conditions for the system (2.1) $$\rho
g_{,\rho}g^{-1}=\pmatrix{0&O(1)\cr0&2\cr},\;
\rho\rightarrow0,\;z\in\Gamma, \eqno(2.4a)$$ $$\hat\Omega_i\rho
g_{,\rho}g^{-1} \hat\Omega_i^{-1}=\pmatrix{2&0\cr
O(1)&0\cr},\;\rho\rightarrow0 \;z\in I_i,\;
\hat\Omega_i=\pmatrix{1&\Omega_i\cr0&1}.\eqno(2.4b)$$ $$\rho
g_{,z}g^{-1}=O(1),\;\rho\rightarrow0,\; z\in R.\eqno(2.4c)$$ where
$\Gamma$ is the symmetry axis consisting of $N+1$ connected
components, $$\Gamma=\bigcup\Gamma_j=R\setminus\bigcup I_i\;\;
(j=1,\ldots,N+1,\; i=1,\ldots,N),\;\; I_i=(z_{2i-1},z_{2i}).$$ The
symbol $O(1)$ denotes uniformly bounded functions on the
corresponding interval. We impose the condition at infinity
$$W=\rho^2 O({1\over r^3}),\;\; X=\rho^2(1+O({1\over r})),
\;\;r=\sqrt{\rho^2+z^2}.\eqno(2.5)$$ Boundary conditions (2.4) and
(2.5) are sufficient to construct solutions of the first group of
Einstein equations (2.1).

We now consider the properties of solutions of system (2.2). Using
(2.3), we can easily verify that for $\rho=0$ and $z\in\Gamma$, we
have $$\partial_z\beta=0,\;\;\beta|_{\Gamma_i}=b_i,\eqno(2.6) $$
where $b_i$ are some constants. The condition for the solution to
be regular on the symmetry axis (${X^{,a}X_{,a}\over
4X}\rightarrow1$ on the symmetry axis) is satisfied if and only if
Eq.(2.3) is supplemented by the additional requirement that
$$b_i=0.\eqno(2.7)$$ For asymptotically flat solutions (those for
which $\beta\rightarrow0$ as $r\rightarrow\infty$) condition (2.7)
is automatically satisfied on the first and the last components of
the symmetry axis but cannot be satisfied on the remaining
components (see \cite{Weinstein90,Weinstein92,Weinstein94,Li91}).
In the general case, the last statement is not yet proved. In this
work, we evaluate the constant $b_i$, but the corresponding
analytic expression is unfortunately too complicated to analyze
the solvability of Eq.(2.7).

The $b_i$ parameters also have their own interpretation: the quantity
$$F_i={1\over4}(e^{-b_i/2}-1)\eqno(2.8)$$
can be considered as the interaction force between the black holes.
Here and what follows, we assume that $b_1=b_{N+1}=0$. We now comment on
the origin of this interpretation. If condition (2.7) is not satisfied,
the curvature tensor becomes singular in the neighborhood of the
corresponding symmetry axis components. We assume that the singular
points are filled with "matter", which is precisely what prevents the
black hole falling on each other. The tension in $z$ direction per
unit surface is ${\rm T}(e_z,e_z)$ where ${\rm T}$ is the energy-momentum tensor
and $e_z$ is the normalized vector parallel to $\partial/\partial z$.
It is natural to define the interaction force of the black holes as the
integral
$$\int_{S_\varepsilon}{\rm T}(e_z,e_z)ds,$$
where $S_\varepsilon$ is the 2-surface coordinatized by
$\rho\phi$(with $\rho\leq\varepsilon)$
and $ds$ is the corresponding area element. Evaluating this integral,
we obtain Eq. (2.8) \cite{Weinstein90}.

To conclude this section, we give a brief derivation of representation
for $b_i$ \cite{Weinstein90,Weinstein92}. We bring Eq.(2.2) to the form
$$\partial_\rho\beta={\rho\over2}\left((\partial_\rho\ln{X\over\rho^2})^2
-(\partial_z\ln{X\over\rho^2})^2+{(\partial_\rho Y)^2-(\partial_z Y)^2
\over X^2}\right),\eqno(2.9a)$$
$$\partial_z\beta=\partial_z\ln X (\rho\partial_\rho\ln X -2)+
{\rho\partial_z Y\partial_\rho Y\over X^2},\eqno(2.9b)$$
where
$$dY={1\over\rho}\ast(XdW-WdX),\;\;(\ast d\rho=dz, \ast dz=-d\rho).$$
Let $C_\varepsilon$ be a curve connecting $\Gamma_{i+1}$ and $\Gamma_i$:
$$\rho=\varepsilon\sin\tau,\;\;z-{z_{2i}+z_{2i-1}\over2}=
-\sqrt{\varepsilon^2+m_i^2}\cos\tau,\;\;0\leq\tau\leq\pi.$$
Then
$$b_{i+1}-b_i=\int_{C_\varepsilon}d\beta.\eqno(2.10)$$
The left-hand side of (2.10) is independent of $\varepsilon$. Taking
$\varepsilon$ to zero and using (2.3) and (2.9), we obtain that
$$b_{i+1}-b_i=-2\left(\ln\hat X(z_{2i})-\ln\hat X(z_{2i-1})\right),
\eqno(2.11)$$
where
$\hat X(z_{2i})=\hat X(m_i,1)$ and $\hat X(z_{2i-1})=\hat X(m_i,-1)$.
In the next section, we use Eq.(2.11) to evaluate $b_i$.

\section{The Riemann problem}

We showed in \cite{Varzugin97} that for every solution of boundary-value
problem (2.4), (2.5) there exists a piecewise analytic matrix $\chi(\omega)$
satisfying the conjugation condition on the imaginary axis
$$\chi_-(\omega)=\chi_+(\omega)
\pmatrix{1&0\cr0&\omega\cr}T(k)\pmatrix{1&0\cr0&1/\omega\cr},
\;\;k=z+(\omega^2-\rho^2)/2\omega\eqno(3.1a)$$
and normalized at infinity by
$$\chi(\omega)\rightarrow I,\;\;
\omega\rightarrow\infty.\eqno(3.1b)$$
The rational matrix $T(k)$ is defined as
$$T(k)=\hat D_{N+1}T_N\hat D_N\ldots T_1\hat D_1,\eqno(3.2)$$
where
$$T_j=\pmatrix{1-{2M_j\over k-z_{2j-1}}&-4M_j\Omega_j\cr
{2L_j\over(k-z_{2j})(k-z_{2j-1})}&1+{2M_j\over k-z_{2j}}\cr},\;\;
\hat D_j=\pmatrix{1&-D_j\cr0&1\cr}.\eqno(3.3)$$
The $M_j$ ¨ $L_j$ parameters have the physical interpretation of the respective
full mass and angular momentum of $j$th black hole and are related by
$$M_j=m_j+2\Omega_j L_j,\;\; 2m_j=z_{2j}-z_{2j-1}.\eqno(3.4)$$
An additional requirement that the matrix $T(k)$ be symmetric leads to
$2N+1$ nonlinear algebraic equations on the parameters $D_j, L_j$ and
$\Omega_j$, which determine $D_j$ and $L_j$ as functions of
$\Omega_j$ and $z_j$. In particular,
$$\sum^{N+1}_{j=1}D_j+\sum^N_{j=1}4\Omega_jM_j=0.\eqno(3.5)$$

Let $d_1=D_1, d_{i+1}=d_i+D_{i+1}+4M_i\Omega_i$ and
$$t_{2j-1}(k)=C_{2j-1}^{-1}\pmatrix{1&0\cr{1\over
2\Omega_j(k-z_{2j-1})} &1\cr}C_{2j-1},\;\; t_{2j}(k)=C_{2j}^{-1}
\pmatrix{1&{\Omega_j\over 2(k-z_{2j})}\cr0&1\cr}C_{2j},$$
where
$$C_{2j-1}=\pmatrix{1&-d_j\cr0&1\cr},\;\;
C_{2j}=\pmatrix{0&\Omega_j\cr -1/\Omega_j&4M_j\cr}C_{2j-1}.$$
Then applying (3.5) and easily verified identity
$$T_j\pmatrix{1&-d_j\cr0&1\cr}t_{2j-1}^{-1}t_{2j}^{-1}=
\pmatrix{1&-d_j-4M_j\Omega_j\cr0&1\cr},$$
we derive that
$$T(k)=t_{2N}(k)t_{2N-1}(k)\ldots t_2(k)t_1(k).\eqno(3.6)$$
We now note that the matrix
$$\pmatrix{1&0\cr0&\omega\cr}t_{2j}(k)t_{2j-1}(k)
\pmatrix{1&0\cr0&1/\omega\cr}$$
has no singularity at $\omega=0$ and tends to identity matrix as
$\omega\rightarrow\infty$,
$$\pmatrix{1&0\cr0&\omega\cr}t_{2j}(k)t_{2j-1}(k)
\pmatrix{1&0\cr0&1/\omega\cr}\Big|_{\omega=0}=
\pmatrix{1&-8M_j(2M_j\Omega_j+d_j)/
\rho^2\cr0&1\cr}.\eqno(3.7)$$
Therefore, the same properties are shared by the matrix
$$\pmatrix{1&0\cr0&\omega\cr}T(k)\pmatrix{1&0\cr0&1/\omega\cr},$$
and Riemann problem (3.1) is correctly posed for any
$D_j, L_j, \Omega_j$ satisfying (3.5).

The only singularities of the conjugation matrix are simple poles at
points
$$\omega^\pm_i=(z_i-z)\pm\sqrt{(z_i-z)^2+\rho^2},\;\;\omega^-_i=
-\rho^2/\omega^+_i.$$
It now follows from Eq.(3.1) that $\chi_\pm(\omega)$ are rational
functions with simple poles at the points $\omega^\pm_i$
($\omega^+_i>0$, $\omega^-_i<0$),
$$\chi_\pm(\omega)=I+\sum_{j=1}^{2N}{A_j^\pm\over
\omega-\omega_j^\pm},\eqno(3.8)$$
with $A_j^\pm$ being independent of $\omega$.
It follows from the unimodularity of $T(k)$ that $\det\chi_\pm(\omega)=1$,
and $\det A_j^\pm=0$. Let $\xi_j$ be the eigenvector of $A_j^-$; then
$$A_j^-\xi_j=0,\;\;A_j^-=a_j \xi_j^\sigma,\;\;\xi_j^\sigma=
(-\xi_j^2,\xi_j^1),\eqno(3.9)$$
with $a_j$ being a column vector. We further note that
$\chi_-(\omega)$ is unimodular if and only if there exists a
constant $v_i$ such that
$$a_i=v_i\lim_{\omega\rightarrow\omega_i^-}\chi_-(\omega)\xi_i,
\eqno(3.10)$$
Equation (3.9) and (3.10) yield a system of linear equations for $a_i$,
$$a_i=v_i\xi_i+v_i\sum_{i\neq k}{\xi^\sigma_k\xi_i\over
\omega_i^--\omega_k^-}a_k.\eqno(3.10a)$$
The parameters $v_i$ and $\xi_i$ can be easily determined from conjugation
matrix.

We define the matrix
$\G_i(k)=t_{i-1}(k)\ldots t_1(k)$
and the vectors
$$c_{2j}=C_{2j}^{-1}\pmatrix{1\cr 0\cr},\;\;c_{2j-1}=C_{2j-1}^{-1}
\pmatrix{0\cr 1\cr}.$$
Then
$$\xi_i=\pmatrix{1&0\cr0&\omega_i^-\cr} \G_i^{-1}(z_i)c_i,\eqno(3.11)$$
and also
$$v_i={\omega_i^- v_i^0\over u_i\omega_i^--v_i^0\xi_i^\sigma
\pmatrix{0&0\cr0&1\cr}\xi_i},
\eqno(3.12)$$
where
$$v_{2j}^0={\Omega_j\over \omega_{2j}^--\omega_{2j}^+},\;\;
v_{2j-1}^0=-{1\over\Omega_j}{1\over \omega_{2j-1}^--\omega_{2j-1}^+}$$
and
$$u_i=1+(\omega_i^--\omega_i^+) v_i^0 c_i^\sigma\partial_k\G_i(z_i)
\G_i^{-1}(z_i)c_i.$$
We also note that
$$\xi_i^\sigma=\omega^-_i c_i^\sigma\G_i(z_i)\pmatrix{1&0\cr0&
1/\omega_i^-\cr}$$

The approach adopted here to solve the Riemann problem is similar to
one used in \cite{Beals84}. As is clear from what follows, to construct
an exact solution to nonlinear
system (2.1) we can take $T$ to be any symmetric matrix. This
method for deriving exact solutions is evidently different from the
method used in \cite{Belinskii-Zakharov78, Belinskii-Zakharov79};
however, it does not lead to new solutions of (2.1).
In particular, the solution that we investigate here is in the class of
$2N$-solitons in Minkowski space background \cite{Belinskii-Zakharov79}.

We now show how the solution of (2.1) can be reconstructed from
the solution of Riemann problem (3.1). Let
$$D_1=\partial_z-{2\omega^2\over\omega^2+\rho^2}\partial_\omega,
\;\;D_2=\partial_\rho+{2\omega\rho\over\omega^2+\rho^2}
\partial_\omega.$$
Because $D_1 k=D_2 k=0$, it follows from (3.1) that
"logarithmic" derivatives $D\Psi\Psi^{-1}$ of the functions
$$\Psi_\pm(\omega)=\chi_\pm(\omega)\pmatrix{1&0\cr0&\omega\cr}$$
have no other singularities in addition to the simple poles at
$\pm i\rho$. Therefore,
$$D_1\Psi_-(\omega)\Psi_-^{-1}(\omega)={-
i\rho\partial_\omega\Psi_-(i\rho)\Psi_-^{-1}(i\rho)\over \omega-i\rho}+
{i\rho\partial_\omega\Psi_-(-i\rho)\Psi_-^{-1}(-i\rho)\over \omega+i\rho}
\eqno(3.13a)$$
$$D_2\Psi_-(\omega)\Psi_-^{-1}(\omega)={
\rho\partial_\omega\Psi_-(i\rho)\Psi_-^{-1}(i\rho)\over \omega-i\rho}+
{\rho\partial_\omega\Psi_-(-i\rho)\Psi_-^{-1}(-i\rho)\over \omega+i\rho}
\eqno(3.13b)$$
We now take into account that $\bar T(\bar\omega)=T(\omega)$ and
define the real matrix
$$g=-\chi_-(0)\pmatrix{1&0\cr0&-\rho^2\cr}.\eqno(3.14)$$
From (3.13) with $\omega=0$, one obtains
$$\partial_\zeta gg^{-1}=\partial_\omega\Psi_-(i\rho)\Psi^{-1}_-(i\rho),
\;\;\bar\partial_\zeta gg^{-1}=\partial_\omega
\Psi_-(-i\rho)\Psi^{-1}_-(-i\rho).\eqno(3.15) $$
The compatibility condition for Eqs.(3.13) is given by system (2.1), and
therefore, Eq.(3.14) is a solution of (2.1).

From here on, we assume that the parameters $d_i$ and $M_i$ are determined
from the requirement that $T(k)$ be symmetric. Then
$$\sum_{i=1}^N M_i(d_i+2\Omega_i M_i)=0,\;\;
(\lim_{k\rightarrow\infty}k(T_{21}(k)-T_{12}(k))=0)\eqno(3.16)$$
and $\chi_-(0)=\chi_+(0)$ (see (3.7)). Using the last identity and the
uniqueness of the solution to the Riemann problem, we obtain the reduced
problem
$$\chi_-(\omega)=-g\tilde\chi_+^{-1}(-\rho^2/\omega)
\pmatrix{1&0\cr0&-1/\rho^2\cr}\;\;
(\tilde T(-\rho^2/\omega)=T(\omega)),\eqno(3.17)$$
where $\tilde T$ denotes the transposed matrix. Taking $\omega$
either to zero or to infinity in (3.17), we see that
$g$ is a symmetric matrix.

We now investigate the properties of solutions of the Riemann
problem as $\rho\rightarrow0$. Let $z\in\Gamma_{m+1}$, then
$$\omega_i^+\rightarrow 2(z_i-z),\;\;\omega_i^-\rightarrow0,
\;\;i\geq 2m+1,\;\;\rho\rightarrow0\eqno(3.18a)$$
$$\omega_i^+\rightarrow0,\;\;\omega_i^-\rightarrow2(z_i-z),
\;\;i\leq 2m,\;\;\rho\rightarrow0\eqno(3.18b)$$ Using the formula
$\omega_i^+=-\rho^2/\omega_i^-$, we can easily verify that all the
coefficients of linear system (3.10) also well defined for
$\rho=0$, and it is therefore natural to assume the existence of
the limit $\lim_{\rho\rightarrow0}a_i$. This assumption ensures
the existence of all the limits used in what follows.

Introduce the matrix
$$\chi_1(\omega)=\chi_-(\omega)\pmatrix{1&0\cr0&\omega}\G_{2m+1}^{-1}
\pmatrix{1&0\cr0&1/\omega}=
\chi_+(\omega)\pmatrix{1&0\cr0&\omega}\Q_{2m+1}
\pmatrix{1&0\cr0&1/\omega} $$
$$=I+\sum_{i=1}^{2m}{B_i^1\over\omega-\omega_i^+}+
\sum_{i=2m+1}^{2N}{B^1_i\over\omega-\omega_i^-},\eqno(3.19)$$
where
$$\Q_i(k)=t_{2N}(k)\ldots t_i(k),\;\;T(k)=\Q_i(k)\G_i(k).\eqno(3.20)$$
We can see from (3.18) that
$$\lim_{\rho\rightarrow0}\chi_1(\omega)=I+{B\over\omega}.\eqno(3.21)$$
Further, let
$$\chi_2(\omega)=\chi_+(\omega)
\pmatrix{1&0\cr0&\omega}\tilde\G_{2m+1}
\pmatrix{1&0\cr0&1/\omega}$$
In view of the identities
$$\pmatrix{1&0\cr0&\omega}\tilde t_{2j-1}\tilde t_{2j}
\pmatrix{1&0\cr0&1/\omega}\Big|_{\omega=0}=I,$$
$$\pmatrix{1&0\cr0&\omega}\tilde t_{2j-1}\tilde t_{2j}
\pmatrix{1&0\cr0&1/\omega}\Big|_{\omega=\infty}=
\pmatrix{1&8M_j(d_j+2M_j\Omega_j)\cr0&1},$$
we have
$$\chi_+(0)=\chi_2(0),\;\;\chi_2(\infty)=
\pmatrix{1&8\sum_{j=1}^{m}M_j(d_j+2M_j\Omega_j)\cr0&1}.
\eqno(3.22)$$

In terms of $\chi_1$ and $\chi_2$, Riemann problem (3.1) can be
rewritten as
$$\chi_2(\omega)=\chi_1(\omega)\pmatrix{1&0\cr0&\omega}\Q^{-1}_{2m+1}
\tilde\G_{2m+1}
\pmatrix{1&0\cr0&1/\omega},\eqno(3.23)$$
and reduced problem (3.17) as
$$\chi_2(\omega)=-g\tilde\chi_1^{-1}(-\rho^2/\omega)
\pmatrix{1&0\cr0&-1/\rho^2\cr}=
B^0+\sum_{i=1}^{2m}{B_i^2\over\omega-\omega_i^-}+
\sum_{i=2m+1}^{2N}{B^2_i\over\omega-\omega_i^+}.\eqno(3.24)$$
Taking $\rho\rightarrow0$ in (3.23) and taking into account (3.21),
we see that
$$\chi_2(\omega)=
B^0+\sum_{i=1}^{2N}{B_i^2\over\omega-2(z_i-z)}
=(I+{B\over\omega})\pmatrix{1&0\cr0&\omega}\gamma_{m+1}(\omega)
\pmatrix{1&0\cr0&1/\omega},\eqno(3.25)$$
where
$$\gamma_{m+1}(\omega)=\Q_{2m+1}^{-1}(z+\omega/2)\tilde\G_{2m+1}(z+
\omega/2).\eqno(3.26)$$
The function $\chi_2(\omega)$ is nonsingular at $\omega=0$. This is
possible only if
$$B=\pmatrix{0&-{\gamma_{m+1}^{12}(0)\over\gamma_{m+1}^{22}(0)}\cr
0&0\cr}$$
Finally,
$$\chi_-(0)=
\pmatrix{1/\gamma^{22}_{m+1}(0)&\left(\gamma_{m+1}^{22}(0)
\partial_\omega\gamma^{12}_{m+1}(0)-\gamma^{12}_{m+1}(0)
\partial_\omega\gamma^{22}_{m+1}(0)\right)/\gamma^{22}_{m+1}(0)\cr
0&\gamma^{22}_{m+1}(0)\cr}\eqno(3.26)$$
Recall that $\chi_-(0)=\chi_+(0)=\chi_2(0)$. We now note that
$\gamma_{m+1}(0)$ is a symmetric real matrix with the unit
determinant and also that $\gamma_{m+1}^{22}(0)\rightarrow1$ as
$z\rightarrow\infty$; therefore,
$$\gamma_{m+1}^{22}(0)>0.\eqno(3.27)$$

Since
$${\hat X\over \lambda^2-m_i^2}\big|_{\Gamma_{i+1}}=\gamma_{i+1}^{22}(0,z),
\;\;{\hat X\over \lambda^2-m_i^2}\big|_{\Gamma_{i}}=\gamma_{i}^{22}(0,z),$$
and $\lambda^2-m_i^2\rightarrow(z-z_{2i})(z-z_{2i-1})$ as
$\rho\rightarrow0$, we find
$$\hat X(z_{2i})=m_i\res_{z_{2i}}\gamma_{i+1}^{22}(0,z),\;\;
\hat X(z_{2i-1})=-m_i\res_{z_{2i-1}}\gamma_{i}^{22}(0,z).\eqno(3.28)$$
It follows from (3.27) that
$$\res_{z_{2i}}\gamma_{i+1}^{22}(0,z)>0,\;\;
\res_{z_{2i-1}}\gamma_{i}^{22}(0,z)<0.$$
Equations (2.11) and (3.28) give the sought-for analytic expression for
the constants $b_i$ and the interaction force (2.8) of black holes.

\section{Interaction force between two black holes}

For two black holes, the symmetry axis has three connected
components. We normalize $\beta$ such that $b_1=b_3=0$, then
$$F={1\over4}\left({\hat X(z_2)\over\hat
X(z_1)}-1\right),\eqno(4.1)$$ and $$\hat
X(z_2)=m_1\res_{z_2}\gamma_2^{22}(0,z),\;\; \hat
X(z_{1})=-m_1\res_{z_1}\gamma_1^{22}(0,z).\eqno(4.2)$$ Expression
(4.1) depends on $d_i, \Omega_i, L_i, M_i$ and $z_i$. These
parameters are not independent and must satisfy the system of
nonlinear algebraic equations
$$\res_{z_i}T_{12}(k)=\res_{z_i}T_{21}(k),\;\; M_i=m_i+2\Omega_i
L_i.\eqno(4.3)$$ It can be easily seen from (3.16) that
$$d_1=-2M_1\Omega_1+M_2d,\;\;d_2=-2M_2\Omega_2-M_1d.\eqno(4.4)$$
Using (4.3) and (4.4), we can eliminate parameters $d_1, d_2,
\Omega_1$ and $\Omega_2$ from the expression for the interaction
force (4.1). Unfortunately, the formula thus obtained is extremely
cumbersome. The situation is greatly simplified in one particular
case, however. Namely, let the black holes rotate in opposite
directions and have the same total and irreducible masses:
$$M_1=M_2=M,\;\; m_1=m_2=m,\;\;\Omega_1=-\Omega_2=\Omega.$$ We set
$$z_1=-m_1,\;z_2=m_1,\;z_3=R-m_2,\;z_4=R+m_2.\eqno(4.5)$$ Then
$$L=2M^2(m+M){R+2M\over R-2M}\Omega,\;\;M=m+2\Omega L,\;\;
L_1=-L_2=L,\; d=0\eqno(4.6)$$ and $$F={M^2\over
R^2-4M^2}.\eqno(4.7)$$ By definition, $R\geq 2m$. Using (4.6), we
can easily show that $2M\leq R$ (where the equality takes place
only in the limiting case of $R=2m$); therefore , $F>0$ and tends
to infinity as $R\rightarrow 2m$. We also observe that the
interaction force in the case under consideration is exactly the
same as for two Schwarzschild black holes with identical masses
(equal to $M$). The interaction force of two nonrotating black
holes ($\Omega_1=\Omega_2=0$) is given by $$F={m_1m_2\over
R^2-(m_1+m_2)^2}.\eqno(4.8)$$

It is easy to see that in the cases we have considered, $F$ tends
to the Newtonian limit as $R\rightarrow\infty$. This result can be
generalized; the estimate $$F={M_1M_2\over R^2}+O({1\over
R^4})\eqno(4.9)$$ holds as $R\rightarrow\infty$. To derive (4.9),
it suffices to use the approximate solution of (4.3):
$$d=-{2L_2M_1+2L_1M_2\over M_1M_2}{1\over R^2}+4\left(L_1+L_2+
{2L_1M_2^2+2L_2M_1^2\over M_1M_2}\right){1\over R^3}+O({1\over
R^4}),$$ $$\Omega_i={L_ia_i\over 2M_i^2(M_i+m_i)}$$ and
$$a_1=1-{4M_2\over R}+{8M_2^2\over R^2}+O({1\over R^3}),\;
a_2=1-{4M_1\over R}+{8M_1^2\over R^2}+O({1\over R^3}).$$

We now discuss the general properties of constraints (4.3) on the
angular velocities and angular momenta. We especially note that
the equality $L_2=\Omega_2=0$ cannot be satisfied for any values of
the remaining parameters with $L_1\neq0$.
This means that if $L_2=0$ and the second black hole does not
contribute to the total angular momentum, an approaching observer would
still see this black hole rotating with a certain angular velocity.
We consider this behavior quite natural. Indeed, the notion of
the angular velocity is related to the behavior of geodesics in the
neighborhood of black hole: every test particle approaching the
black hole is involved in its rotation, and the angular velocity
of this rotation becomes equal to the angular velocity of the
black hole when the particle reaches the event horizon.
That the angular velocity is nonzero is therefore determined by
the fact that the first black hole involves the second one in its
rotation. This explanation corresponds to the behavior
of $\Omega_2$ at large distances: if $L_2=0$ then $\Omega_2=O(1/R^3)$.

We return to the exact solution (4.6). The foregoing discussion
allows us to take the angular momenta of each black hole as the
independent parameters. We then have $\Omega\rightarrow0$ and
$M\rightarrow m$ as $R\rightarrow 2m$ which means that as black holes
approach each other, they loose the full mass and "slow down" each
others rotation.

Evidently, the realistic solution for the configuration of two
black holes cannot be stationary. As we have seen however, the
stationary solutions admits a sufficiently realistic physical
interpretation and demonstrate, at least qualitatively, the
effects caused by the interaction of black holes. In our opinion,
this is because the Einstein equations determine not only the
gravitational field but also the law of motion of material bodies
\cite{Fock39}.

This work was supported by RFBR grant No. 96-01-00548.

\end{document}